\documentclass[12pt]{article}
\pdfoutput=1
\usepackage{amsmath,amssymb,amsfonts,epsfig,graphicx,euscript}%
\usepackage[pdftex,bookmarks,bookmarksnumbered,linktocpage,pdfstartview=FitH]{hyperref}
\hypersetup{colorlinks,%
citecolor=red,%
filecolor=blue,%
linkcolor=blue,%
urlcolor=blue,%
pdftex}
\usepackage[all]{hypcap}
\numberwithin{equation}{section}
\usepackage{cite}
\newcommand{\bse}{\begin{subequations}}
\newcommand{\ese}{\end{subequations}}
\newcommand{\be}{\begin{equation}}
\newcommand{\ee}{\end{equation}}
\newcommand{\bea}{\begin{eqnarray}}
\newcommand{\eea}{\end{eqnarray}}
\newcommand{\ba}{\begin{array}}
\newcommand{\ea}{\end{array}}
\newcommand{\h}{\frac{1}{2}}

\begin{document}
\hfill%
\vspace{0.5cm}
\begin{center}
{ \Large{\textbf{Thermal Fluctuations and Meson Melting: A Holographic Approach}}} 
\vspace*{1.2cm}
\begin{center}
{\bf M. Ali-Akbari$^{a,}$\footnote{m$\_$aliakbari@sbu.ac.ir}, Z. Rezaei$^{b,d,}$\footnote{z.rezaei@aut.ac.ir},
A. Vahedi$^{c,}$}\footnote{vahedi@ipm.ir}\\%
\vspace*{0.3cm}
{\it {${}^a$Department of Physics, Shahid Beheshti University G.C., Evin, Tehran 19839, Iran}}\\
{\it {${}^b$Department of Physics, Tafresh University, 39518-79611, Tafresh, Iran}}\\
{\it {${}^c$Department of physics, Kharazmi University, P.O.Box 31979-37551, Tehran, Iran}}\\
{\it {${}^d$School of Particles and Accelerators, Institute for Research in Fundamental Sciences (IPM),
P.O.Box 19395-5531, Tehran, Iran}}  \\

\vspace*{0.8cm}
\end{center}
\end{center}

\bigskip
\begin{center}
\textbf{Abstract}
\end{center}
We use gauge/gravity duality to investigate the effect of thermal
fluctuations on the dissociation of the quarkonium mesons in strongly
coupled $(3+1)$-dimensional gauge theories.
The purpose of this paper is to introduce \textit{a new approach}
to study the instability and probable first order phase transition of a probe
D7-brane in the dual gravity theory. We explicitly
show that for the Minkowski embeddings with their tips close to the
horizon in the background, the long wavelength thermal fluctuations
lead to an imaginary term in their action signaling an instability
in the system. Due to this instability, a phase transition is
expected. On the gauge theory side, it indicates that the quarkonium
mesons are not stable and dissociate in the plasma. Identifying the
imaginary part of the probe barne action with the thermal width of
the mesons, we observe that the thermal width increases as one
decreases the mass of the quarks. Also keeping the mass fixed,
thermal width increases by temperature as expected. We will also
investigate the effect of the magnetic field on the mass and the
thermal width.

\newpage

\tableofcontents

\section{Introduction and Results}
One of the main phenomena observed in Quark-Gluon Plasma (QGP),
which is experimentally produced at Relativistic Heavy Ion Collider
(RHIC) by colliding two heavy nuclei (such as gold or lead) \cite{Brambilla:2004wf,
CasalderreySolana:2011us}, is
quarkonium meson dissociation (or melting) \cite{Alessandro:2004ap}. Quarkonia survive as bound states in the
QGP up to a dissociation temperature $T_{diss}$. If the produced QGP
is hot enough, $T>T_{diss}$, quarkonia will dissociate in the plasma. In
fact the dissociation temperature is realized as a temperature in
which meson completely ceases to exist. Since the QGP is a strongly
coupled plasma, computing the dissociation temperature
perturbatively is not trustworthy and therefore a non-perturbative
framework such as gauge/gravity duality will be useful to describe
various aspects of the QGP including the dissociation temperature. Of course it is worth mentioning that in thermal QCD in the perturbative high temperature sector, the thermal width of the quarkonium bound state has been analyzed in great detail for instance in \cite{Burnier:2007qm}.

Gauge/gravity duality is a duality between a strongly coupled non-abelian gauge
theory and a weakly coupled theory of gravity in a higher dimensional spacetime where the gauge theory lives on the boundary of the spacetime \cite{Maldacena, CasalderreySolana:2011us}. In fact it is a really helpful duality since by using the gravity dual the analytic calculation for the strongly coupled gauge theory, which is not tractable already, can be done.  In particular, the strongly coupled ${\cal{N}}=4$ super Yang-Mills (SYM) theory with gauge group $SU(N)$ in (3+1) dimensions is dual to type IIb supergravity on $AdS_5\times S^5$. This duality has found many generalizations and specifically it has been successful in describing gauge
theories at finite temperature \cite{Witten:1998qj}. In short a thermal SYM theory
corresponds to the supergravity in an AdS-Schwarzschild black hole background where the
temperature of the gauge theory is identified with the Hawking temperature of the black hole.

In the context of gauge/gravity duality, adding fundamental matter (quark) to the strongly
coupled SYM theory which lives on the boundary of AdS space is done by introducing the branes
into the dual gravity theory that lives in the bulk. This has been proposed in \cite{Karch:2002sh} where $N_f$ D7-branes are added to the $AdS_5\times S^5$ background in the probe limit that means $\frac{N_f}{N}\ll 1$. This will introduce a matter hypermultiplet in the fundamental representation of the gauge group $SU(N)$ to the SYM theory. The hypermultiplet describes the dynamical quarks living in 3+1 dimensions and the fluctuations of the probe brane explain the meson spectrum of the field theory \cite{CasalderreySolana:2011us, Erdmenger:2007cm}.

The embeddings of the probe brane in the AdS-Schwarzschild black hole background are
classified into three types \cite{Mateos:2006nu}. The Minkowski embeddings (MEs) are those in which the probe brane closes off smoothly  above the horizon of the background black hole or equivalently the induced metric on the probe brane has no horizon. For sufficiently large $m_q/T$, where $m_q$ and $T$ are the bare mass of the quark and temperature of the SYM theory respectively, these embeddings are energetically favorable. In the black hole embedding (BE) case, the induced metric on the brane has an event horizon inherited from the horizon of the black hole living in the background. Between these two types of embeddings, there is a critical embedding for which the tip of the brane touches the horizon at a point.

A (discontinuous) first order phase transition between ME and BE is
well known and has been studied in the literature. It is assumed that the mentioned
phase transition happens because $c$, which is the expectation value
of the dual operator to the mass, is a multi-valued function of
$m_q$ \cite{Mateos:2006nu, Mateos:2007vn}. However, we present an explicit calculation for this phase transition from ME to BE which is based on considering
the effect of \textit{thermal fluctuations} on the shape of the brane to make its action imaginary. It will be interestingly seen that applying thermal fluctuations' method previous results of phase transition are reproduced. Moreover this method has the following advantages: 1) This setup is simple and tractable. 2) The results are general and can be applied to various backgrounds or equivalently it provides us with the possibility of describing different strongly coupled SYM theories. 3) Utilizing this setup, we can discuss how thermal fluctuations affect the system. 4) The thermal width for the mesons can be easily defined which is consistent with our physical intuition.

This branes' shape transition corresponds to the dissociation of the heavy quarkonium mesons on the gauge theory side
\cite{CasalderreySolana:2011us, Mateos:2006nu, Mateos:2007vn}. The
MEs are more favorable in the low temperature limit, \textit{i.e.}
$m_q/T \gg 1$ or equivalently $T<T_{diss}$. On the MEs the
quarkonium mesons are bounded and their spectrum is discrete and
gapped. On the other hand, for $T>T_{diss}$ where BEs are more
favorable, the quarkonium mesons are unstable and therefore they disappear in the QGP \cite{Hoyos:2006gb}. Using the holographic
method, the numerical calculations show that the dissociation
temperature is of order $ m_q/\sqrt{\lambda}$ \cite{Mateos:2006nu,
Mateos:2007vn} where $\lambda$ is 't Hooft coupling constant.

Since the above phase transition occurs before the critical embedding is reached, it
is expected that MEs which are the near-critical embeddings turn out to be unstable \cite{Mateos:2007vn}\footnote{In \cite{Mateos:2007vn}, by computing the meson spectrum for the near-critical MEs, it was shown that the tachyonic modes appear in the mass spectrum \textit{signaling} that these embedings are expected to be unstable from thermodynamic considerations. Applying a new approach, as we will see in section 3, we acquire an imaginary term contributing in the probe brane action due to the \textit{long wavelength thermal fluctuations}. As a result, for the first time we show that this family of the thermal fluctuations is responsible for the instability of these embeddings and we hence confirm  the mentioned expectation in \cite{Mateos:2007vn}. It is therefore worth noticing that we put emphasis on the imaginary part of the probe action instead of tachyonic modes of the meson. }.
This expectation is confirmed by explicit calculation in this paper and our main result is that the long wavelength thermal fluctuations\footnote{This family of fluctuations about a classical string configuration has been applied in \cite{Finazzo:2013rqy} to compute
the decay rate of the large spin meson. The same fluctuations in the anisotropic
background \cite{Fadafan:2013bva}, for the moving quarkonium  mesons \cite{Ali-Akbari:2014vpa}
and in the presence of finite 't Hooft coupling correction \cite{Fadafan:2013coa} have been discussed. In this paper we have applied the same fluctuations about the shape of the probe brane. }
lead to an imaginary term in the action of the brane for the near-critical embeddings. This imaginary term indicates an instability in the system and therefore a phase transition
(from ME to BE) happens. As it is well-known, this situation corresponds to meson dissociation in the dual
gauge theory and so we identify the imaginary part of the action with the thermal
width that meson acquires. The behavior of the thermal width with respect to the quark
mass will also be studied. For masses larger than a critical mass, $m_c$, the thermal
width is zero and for masses below this critical value thermal width will increase.
It will be seen that thermal width increases with temperature for a fixed value of mass
as expected \cite{CasalderreySolana:2011us}.

We will also investigate the effect of the magnetic field on the mass and the thermal width.
The action of the probe brane, which describes the dynamics of the brane,
will find an imaginary term in the presence of the magnetic field, too, and the behavior of
thermal width with mass is similar to the zero magnetic field case. In the absence of the
magnetic field the critical mass increases linearly with the increase in temperature while
this dependence is not linear when the magnetic field is non-zero. In the presence of the
magnetic field there is a critical temperature above it the critical mass exists and below
it the critical mass is zero and therefore MEs are stable. It will be shown that in a
constant temperature the magnetic field leads to the decrease of the critical mass.

This paper has the following structure. In section 2 we review MEs in a general
background with a magnetic field turned on. Then we choose
$AdS_5\times S^5$ background in order to be specific with the
asymptotic behavior of the probe brane. In section 3 we study the
effect of thermal fluctuations about the ME on the action of the
brane to obtain an imaginary part. Section 4 is devoted to the
thermal width acquired by quarkonium meson that is the holographic
dual of the imaginary part of the brane action. How the
magnetic field affects the thermal width
is also discussed in this section. Finally, summary and results are
presented in section 5.

\section{Minkowski Embeddings in the Near Horizon Region}
In this section we are interested in finding the profile of the probe brane in a specific background. This profile specifies by scalar fields transverse to the probe brane and depends generally on the background as well as on which components of the gauge field living on the brane have been turned on. In the AdS-Schwarzschild black hole background, for zero and nonzero magnetic field, we find MEs and show that for a given value of the mass of the quarks there are two solutions in the near horizon region. We then discuss how one can determine which of them is the physical solution. In fact, lower energy condition helps us to find the physical configuration.

Let us consider a general form for the background metric as %
\be\begin{split}\label{background}
 ds^2=&-g_{tt}dt^2+g_{xx}dx^2+g_{yy}dy^2+g_{zz}dz^2+ g_{rr}dr^2 \cr
 &+g_{ss}ds^2_{S^3}+g_{RR}dR^2+g_{\varphi\varphi}d\varphi^2,
\end{split}\ee
which is asymptotically $AdS_5\times S^5$. The gauge theory lives in $t, x, y$ and $z$ directions. The $r$ denotes the radial direction and the boundary is located at infinity. It is also assumed that all the components of the metric are functions of the radial direction. Moreover the above metric describes a black hole geometry that means $g_{tt}=g^{rr}=0$ at the horizon.

The effective action of the D7-brane in a general background is given by
\be\begin{split} %
 S_{{\rm{DBI}}}=-\ \tau_{7}\int d^{8}\xi\
 e^{-\phi}\sqrt{-\det(G_{ab}+2\pi\alpha'F_{ab})} ,\cr
\end{split}\ee %
where $\tau_7^{-1}=(2\pi)^7l_s^8g_s$ is the D7-brane tension and $\xi^a$ are worldvolume coordinates. $G_{ab}=g_{MN}\partial_a X^M\partial_b X^N$ is the induced metric on the brane and $g_{MN}$ has been introduced in \eqref{background}. $F_{ab}$ is the field strength of the gauge field living on the brane. We use static gauge meaning that the brane is extended along $t, x, y, z, r$ and $S^3$. Note also that the Dilaton field, $\phi$, can be non-trivial for the background we have considered in \eqref{background}.

We specify the position of the brane in the $R\varphi$-plane as $R(r)$ and $\varphi=\varphi_0$ where $\varphi_0$ is a constant. Because of the translational and rotational symmetry in $\{t,x,y,z\}$ and $S^3$ directions, $R$ depends only on the radial direction and gives the shape of the brane. Hence the induced metric of the probe D7-brane becomes
\be\begin{split}
 ds^2_{brane}=&-g_{tt} dt^2+g_{xx}dx^2+g_{yy}dy^2+g_{zz}dz^2\cr %
 &+dr^2(g_{rr}+g_{RR}\dot{R}^2) +g_{ss}ds^2_{S^3},
\end{split}\ee
where $\dot{R}=\frac{dR(r)}{dr}$. Since we would like to study the effect of the constant magnetic field on the imaginary part of the action, we also include
\be
 B_z=F_{xy}.
\ee
In the end the action of the brane is given by $S=-\tau_7 {\rm{V}}_3{\rm{V}}(S^3) \int dt dr {\cal{L}}$ where
\be\label{lagrangian} %
 {\cal{L}}=e^{-\phi} \left(g_{ss}^3 g_{rr}\right)^{1/2}\sqrt{P+Q \dot{R}^2},
\ee %
and
\bse\begin{align}
 \label{MM} P&=g_{tt}g_{zz}\big(g_{xx}g_{yy}+(2\pi\alpha')^2B_z^2\big), \\
 Q&=\frac{g_{RR}}{g_{rr}} P.
\end{align}\ese
From this final form of the action, we can derive the equation of motion for the $R(r)$. Since it is not illuminating, we do not write it here explicitly. In addition this equation is complicated to be solved analytically and therefore we use numerical method to solve it. In order to find a solution for the equation of motion we need two boundary conditions. If we call this solution $R_{0}$, the first boundary condition is $R_0(0)=R_*$. The regularity condition of the brane is the second one which is given by $\dot{R}_0(0)=0$. Applying the above conditions, various solutions to the equation of motion can be found. This type of solutions, $R_0(r)$, are called MEs. Various aspects of the ME has been studied in the literature, for instance see \cite{CasalderreySolana:2011us, Erdmenger:2007cm, ahmad}.

According to the AdS/CFT dictionary, for different fields leading and sub-leading terms in near boundary asymptotic expansion define a source for the dual operator and its expectation value,
respectively. For $R_0(r)$, in the unit in which the radius of $AdS_5$(or $S^5$) is equal to one, this expansion is
\be\label{asymptotic} %
 R_0(r)=m+\frac{c}{r^2}+ \dots,
\ee %
where the leading term is proportional to the mass of the fundamental matter and the
sub-leading term is proportional to $\langle {\cal{O}}_m\rangle$ where ${\cal{O}}_m$ is the dual
operator to mass and according to \cite{Hoyos:2011us}, we have %
\be\begin{split}
 m_q &=\sqrt{\frac{\lambda}{2\pi}}~ m, \cr
 \langle {\cal{O}}_m\rangle &=-\frac{\sqrt{\lambda}}{4\pi^3}N c.
\end{split}\ee

\begin{figure}[ht]
\begin{center}
\includegraphics[width=2.6 in]{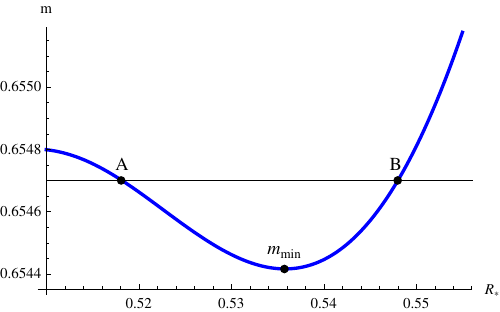}
\caption{The mass of the fundamental matter as a function of $R_*$ at the fixed temperature $T= 0.22733$. The minimum value of mass is given by $m_{min}=0.654416\ (R_{* min}=0.53575)$. }
\label{mRs}
\end{center}
\end{figure}

In order to study the asymptotic behavior of the probe brane shape, we need to be more specific with the background so we consider $AdS_5\times S^5$ background metric produced by a near horizon geometry of $N$ D3-branes,  %
\be\label{ads}\begin{split} %
g_{tt}&=\rho^2\frac{(1-\frac{\rho_h^4}{\rho^4})^2}{1 + \frac{\rho_h^4}{\rho^4}},\ g_{xx}=g_{yy}=g_{zz}=\rho^2(1 + \frac{\rho_h^4}{\rho^4}), \cr
g_{rr}&=g_{RR}=\frac{g_{ss}}{r^2}=\frac{g_{\varphi\varphi}}{R^2}=\frac{1}{\rho^2},
\end{split}\ee %
where $\rho^2=r^2+R^2$ and the Dilaton field is a constant. $\rho_h$ denotes the location of the
horizon and accordingly the temperature of the black hole is $T=\frac{\sqrt{2}\rho_h}{\pi}$ which
is identified with the temperature of the SYM theory. As it was already mentioned in the previous
paragraph, one can solve the equation of motion for $R(r)$ and determine the mass of the fundamental
 matter from the asymptotic value of this solution as is shown in (\ref{asymptotic}). Setting the
 magnetic field to zero, the mass in terms of $R_*$ has been plotted in Fig. \ref{mRs}. Note that for the MEs $R_*$ is always larger than $\rho_h$.

Our numerical calculation shows that there is no one to one
correspondence between mass and $R_*$ in the near horizon
region\footnote{These multivalued solutions in the near horizon region can be clearly seen when one plots the mass of the quarks in terms of the condensation \cite{Mateos:2006nu}.}. In fact for a fixed value of mass there are
two possible solutions (see points A and B in the Fig. \ref{mRs})
and the question is that between these two points which probe brane
configuration is more stable. Since the system is in equilibrium,
the configuration with lower energy is more stable.
To compare the energies one needs to find the Hamiltonian which in our case is given by %
\be\label{hamiltoni} %
 E=\int_{r_*}^{\infty} dr {\cal{H}}-{\cal{L}}_{\rm{cont}},
\ee %
where ${\cal{H}}=-{\cal{L}}$. ${\cal{L}}_{\rm{cont}}=L_1+L_2+L_3$
represents the counterterms that renormalize the divergence at the
boundary and $L_1$, $L_2$ and $L_3$ are explicitly introduced in
\cite{Karch:2007pd, Hoyos:2011us}. It is found from our numerical
analysis that the configuration corresponding to point B has lower
energy and as a result the solution with larger $R_*$ is the
physical one. Similarly other points on the right part of the curve
has lower energy in comparison with the corresponding points on the
left side that have the same mass. Therefore for the physical
configurations, $R_*$ is always larger than $R_{*min}$ corresponding
to $m_{min}$.
\begin{figure}[ht]
\begin{center}
\includegraphics[width=2.6 in]{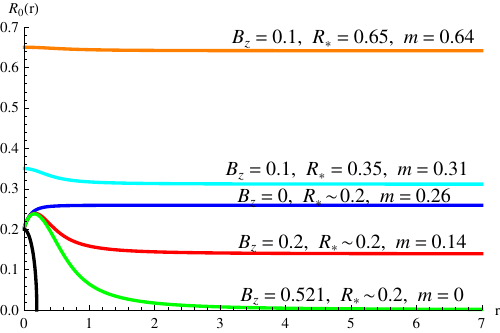}
\caption{Behaviour of the function $R_0(r)$ at the fixed temperature $T= 0.0900316$ for different values of the magnetic field.}
\label{RrB}
\end{center}
\end{figure}%

When the magnetic field is non-zero, as it is evidently seen from
Fig. \ref{RrB}, the value of the mass decreases as the magnetic
field increases. Therefore, at fixed temperature and $R_{\ast}$ an
upper limit exists for the magnetic field. For larger values of the
magnetic field the mass of the quark becomes negative and the
corresponding configurations are not physical \cite{Babington:2003vm}. Moreover, we would
like to emphasis that when $R_* \gg \rho_h\sim T$ the embedding
solution $R_0(r)$ is almost constant (cyan and orange curves in Fig.
\ref{RrB}). However, in the
 near horizon region where $R_*\gtrsim \rho_h$, meaning that the tip of the brane
 is close to the horizon, non-trivial solutions exist (blue, red and green curves in Fig. \ref{RrB}).

In the presence of the magnetic field, although the energy of the system
is still given by \eqref{hamiltoni},  the Lagrangian of the counterterms has now an extra term  %
\be %
{\cal{L}}_{\rm{cont}}=L_1+L_2+L_3+L_f,
\ee %
where $L_f$ has been also introduced in \cite{Karch:2007pd,
Hoyos:2011us}, explicitly. Our results show that, even in the
presence of the magnetic field, for fixed values of mass the
solutions with larger $R_*$ (i.e. $R_*>{R_*}_{min}$) have lower
energy and therefore are more stable. Different aspects of the probe
brane in the presence of the magnetic field have been studied in the
literature \cite{Karch:2007pd, Hoyos:2011us, ahmad}.

\section{Thermal Fluctuations and Imaginary Part of the Action}
In this section we are interested in studying the effect of a
certain class of fluctuations about the ME on
the action of the brane. Due to these fluctuations, an imaginary
term induces on the action of the probe brane. We will find the
general expression for the imaginary part of the action in terms of metric components and external magnetic field.

To compute the effect of the fluctuations on the probe
brane action, the starting point is to define the fluctuations
as\footnote{Notice that, in contrast to mesons denoting by $\delta R(r)$ in \cite{Mateos:2007vn}, here $\delta R(r)$ represents the long wavelength thermal fluctuations.}
\be
 R_0(r)\rightarrow R_0(r)+\delta R(r),
\ee
where $R_0(r)$ and $\delta R(r)$ denote a ME and
fluctuations about it, respectively. $\delta R(r)$ is taken to be
small and we also impose the long wavelength condition on the
fluctuations meaning $\frac{d(\delta R)}{dr}\rightarrow 0$. As a result it is acceptable to expand the ME
around the $r=0$ because the fluctuations have the most
contribution around this point. Regarding the second boundary
condition ($\dot{R}_0(0)=0$), we have
\be\label{RM} %
 R_0(r)=R_*+\frac{1}{2!}\delta r^2\ddot{R}_0(0)+\dots,
\ee %
and, using the above equation, it is easy to find %
\be\label{M}\begin{split}
 P=P(R_*)&+\frac{1}{2!}\delta r^2\ddot{R}_0(0)P'(R_*)+P'(R_*)\delta R(0)\cr
 &+\frac{1}{2}P''(R_*)\delta R(0)^2+\dots,
\end{split}\ee
where $'=\frac{d}{dR}$ and $P$ has been defined in \eqref{MM}. Similar expansion with \eqref{M} can be
written for $Q$. After substituting \eqref{M} in the term under the
square root in \eqref{lagrangian} we have
\be\label{Simaginary} %
 \sqrt{C_1 \delta r^2+C_2},
\ee %
where
\bse\label{ahmad}\begin{align}
 \label{C1} C_{1}&=\frac {\ddot{R}_0(0)}{2}\left(2\ddot{R}_0(0)Q(R_*)+P'(R_*)\right),\\
 \label{C02} C_{2}&=P(R_*)+\delta R(0) P'(R_*) +\frac{1}{2}\delta R(0)^2 P''(R_*),
\end{align}\ese
up to second order in $\delta r$ and $\delta R$.

Since we are working in the classical gravity regime, the saddle point approximation for $\delta R$ can be used to find the main contribution of the thermal fluctuations to the action. Therefore, the value of \eqref{C02} can be analytically computed in this approximation. Using \eqref{Simaginary} and \eqref{C02} it is easy to obtain
\be\label{deltaR}
 \delta R(0)=-\frac{P'(R_*)}{P''(R_*)},
\ee
and by substituting \eqref{deltaR} into \eqref{ahmad}, we finally have
\be\label{C2} %
 C_{2}=P(R_*)-\frac{P'(R_*)^2}{2P''(R_*)}.
\ee %

The imaginary part of the action, if any, arises form \eqref{Simaginary}. We can not discuss about the sign of $C_1$ and $C_2$ generally. Therefore, regarding our numerical calculation, we
assume that $C_1>0$ and $C_2<0$ in a specific region of $R_*$. Hence the expression under the square root in \eqref{Simaginary} is negative in the following region
\be\begin{split} %
 0<\delta r <\delta r_c &=\sqrt{\frac{-C_2}{C_1}} \cr
 &=\left[\frac{-2\Big(2P''(R_*)P(R_*)-P'(R_*)^2\Big)}{\ddot{R}_0(0)P''(R_*)\left(2\ddot{R}_0(0)Q(R_*)+P'(R_*)\right)}\right]^{\h},
\end{split}\ee%
and in this region, in the lowest order, the imaginary part is consequently given by %
\be \label{act3}
{\rm{Im}}S \sim \int_0^{\delta r_c} d(\delta r) \ e^{-\phi(R_*)}(g_{ss}^3g_{rr})^\h_{R=R_*} \sqrt{C_1 \delta r^2+C_2}.
\ee
Equations \eqref{ahmad}-\eqref{act3} are quite general and applicable
to any arbitrary background. In a specified background, for instance
$AdS_5\times S^5$, for each value of $R_*$ one can find a ME,
$R_0(r)$. Using $R_*$ and the second derivative of $R_0(r)$ at $r=0$, the
functions $C_1$, $C_2$ and consequently, from \eqref{act3},
${\rm{Im}}S$ are determined specifically.

\begin{figure}[ht]
\begin{center}
\includegraphics[width=2.6 in]{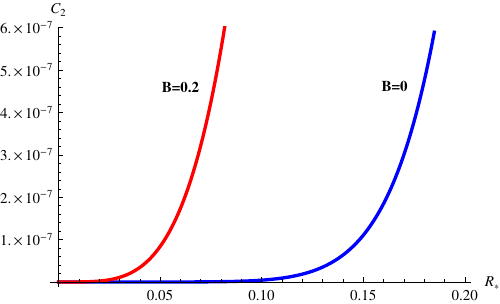}
\caption{Dependence of $C_2$ on $R_*$ at zero temperature in the presence and absence of the magnetic field.} \label{C2RsTzero}
\end{center}
\end{figure}

It is worth noting that $C_2$ must be negative in the near horizon region in order for the action to gain an imaginary part. Our numerical computations for AdS background, \eqref{ads}, show that this requirement is established just when the background is at "finite temperature". It is obvious from Fig. \ref{C2RsTzero} that $C_2\geq 0$ at zero temperature which means that the action will not develop an imaginary part. This is the reason that we use the expression of "thermal fluctuations" for the mentioned class of fluctuations.

\subsection{More on Thermal Fluctuation}
\begin{figure}[ht]
\begin{center}
\includegraphics[width=2.6 in]{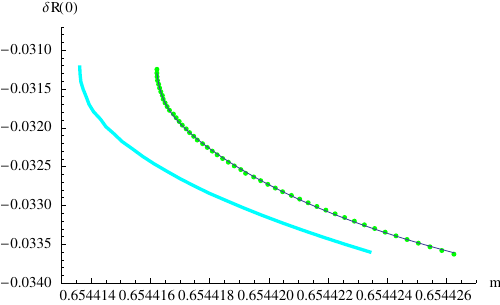}
\includegraphics[width=2.6 in]{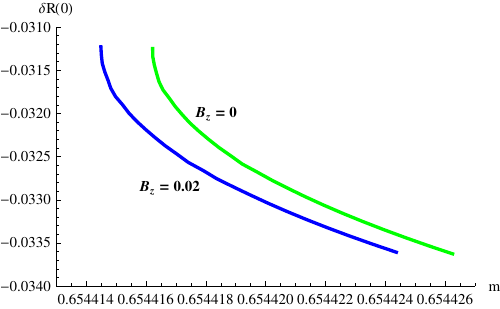}
\caption{The above figures have been plotted between critical mass(see after \eqref{newmass} for definition) and $m_{min}$.\newline
Left: Thermal fluctuations in terms of mass for $T=0.22733$(green) and $T=0.227329$(cyan). The blue curve shows the fitted function \eqref{deltam}. \newline
Right: Thermal fluctuations in terms of mass for non-zero magnetic field and $T=0.22733$.  } \label{delta}
\end{center}
\end{figure}

Before giving a prescription for calculating thermal width, we would
like to mention a few points about thermal fluctuations which are
compatible with our physical intuition (see Fig. \ref{delta}).
\begin{itemize}
\item As it was obtained in \eqref{deltaR}, in our calculations the fluctuations can be written in terms of the metric components and the external magnetic field. In $AdS_5\times S^5$ background our numerical computations display that the value of the thermal fluctuations is negative (in a region of $R_*$ with $C_2<0$).
\item Raising the temperature the fluctuations, needed to make the brane configuration unstable, become smaller at fixed mass. The reason is simply that the tip of the probe brane becomes closer to the horizon for higher temperature.
\item At fixed temperature, in order to have an imaginary term in the brane action, the absolute value of the fluctuations increases for larger values of mass. It seems reasonable because for larger masses the tip of the probe brane goes further away from the horizon.
\item If we attempt to gain an adequate function describing the thermal fluctuations in terms of mass, the best fit we find is %
\be\label{deltam} %
\delta R(0)=\frac{\alpha_1+\alpha_2 m^2+\alpha_3 m^4}{\alpha_4+\alpha_5 m^2+\alpha_6 m^4},
\ee %
where $\alpha$s are temperature dependent coefficients. Using
various computational software programs, such as Mathematica, the
coefficients can be found and the resulting function has been shown
by blue curve in Fig. \ref{delta}(left). Since the values of
$\alpha$s are not illuminating, we did not write them here
explicitly.
\item In Fig. \ref{delta}(right) the fluctuations in terms of mass has been plotted in the presence of the external magnetic field. It is obvious that for non-zero magnetic field the fluctuations have to increase in order to make the configuration unstable. This happens because the magnetic field increases the effective tension of the probe brane and the brane resists more against the deformation.
\end{itemize}

\section{Holographic Thermal Width}
From the gauge/gravity duality point of view, the dissociation of the mesons
is identified with a first order phase transition between a ME and a
BE. Holographic calculation shows that the dissociation temperature
$T_{diss}$ is proportional to the bare mass of the quark $T_{diss}\sim m_q/\sqrt{\lambda}$ \cite{Mateos:2006nu,
CasalderreySolana:2011us}. Notice that the phase transition of the
probe brane's shape means that some of the MEs turn out to be
unstable in the near horizon region.

As it was discussed in \eqref{asymptotic} for the
MEs, the mass of the fundamental matter is proportional to the asymptotic value of the $R_0(r)$, %
\be\label{newmass} %
 m=\lim_{r\rightarrow\infty}R_0(r).
\ee %
In the following subsections at fixed temperature we will find a
critical mass $m_c$ where for available mass smaller than $m_c$ an
imaginary term contributes to the action of the probe branes. In
other words, it indicates that the corresponding MEs are not stable
for $m_{min}<m<m_c$ and a phase transition (first order) from MEs to
BEs may occur ($m_{min}$ was introduced in Fig. \ref{mRs}) . In the gauge theory when the mass of the quark is
less than $m_c$ the quarkonium mesons are unstable. Such mesons are
disappeared (melted) into the plasma and a thermal quark is created
from the medium. Therefore one can define a thermal width for the
quarkonia. We consider the imaginary
part of the action of the probe brane proportional to the thermal
width of the quarkonium mesons in the gauge theory \footnote{In \cite{Hashimoto:2013mua} the imaginary part of the DBI action is identified with the vacuum decay rate.

In a weak-coupling effective field theory the thermal width and quarkonium meson dissociation have been also studied,  for example see \cite{Brambilla:2013dpa}. At strong coupling, lattice studies have shown that the potential may have  an imaginary part \cite{Rothkopf:2011db}.}
\be\label{ImS} %
 \Gamma\sim{\rm{Im}}S.
\ee%
Obtaining the imaginary part of the action of the probe brane from
\eqref{act3}, we aim to compute the thermal width of quarkonia in
the ${{\cal{N}}=4}$ gauge theory. In the following subsections we
like to study the behavior of the thermal width in the absence and in
the presence of a magnetic field.

\subsection{Zero Magnetic Field}
\begin{figure}[ht]
\begin{center}
\includegraphics[width=2.6 in]{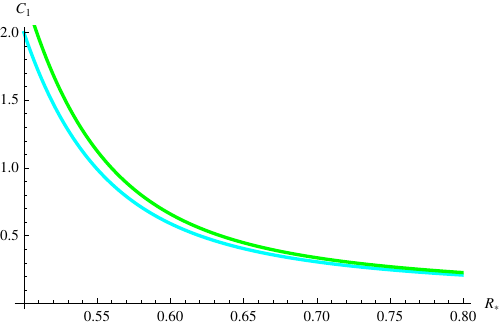}
\includegraphics[width=2.6 in]{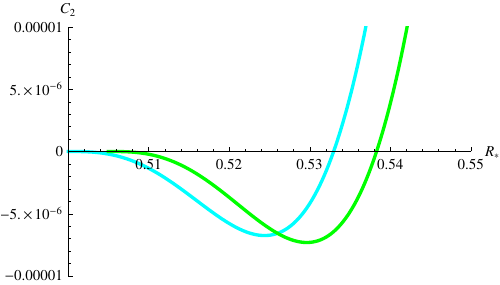}
\caption{$C_1$ in terms of $R_*$ (Left) and $C_2$ in terms of $R_*$
(Right) for two different temperatures $T=0.22508$ and $0.22733$
corresponding to cyan and green curves, respectively. } \label{C2m}
\end{center}
\end{figure}

\begin{figure}[ht]
\begin{center}
\includegraphics[width=2.6 in]{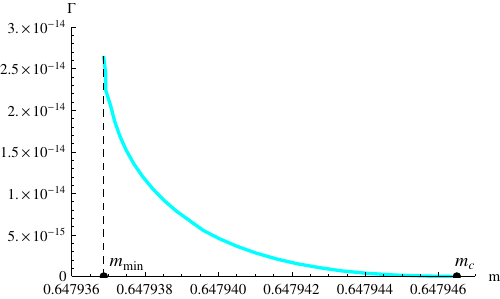}
\includegraphics[width=2.6 in]{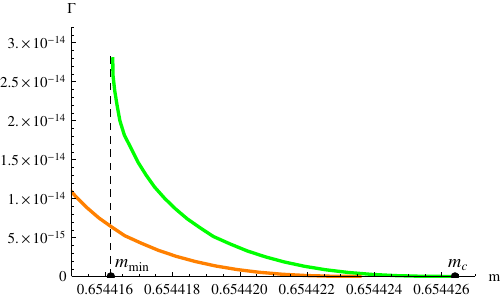}
\caption{Dependence of the thermal width on the mass. Left:
$T=0.22508$. Right:  $T=0.22733$ and $0.227329$ corresponding to green and
orange curves, respectively. } \label{Gm}
\end{center}
\end{figure}

We start with the metric \eqref{ads} corresponding to
AdS-Schwarzschild background and we also set the magnetic field to
zero. The dependence of the functions $C_1$ and $C_2$ on $R_*$ and $\rho_h$ is given by
\be\label{newC}\begin{split} %
C_1&=\frac{\ddot{R}_0(0)(R_*^8-\rho_h^8)\left(4(R_*^8+\rho_h^8)+\ddot{R}_0(0)R_*(R_*^8-\rho_h^8)\right)}{R_*^9},\cr
C_2&=\frac{(3R_*^8-5\rho_h^8)(R_*^8-\rho_h^8)^3}{R_*^8(7R_*^{16}+9\rho_h^{16})},
\end{split}\ee %
and for fixed temperatures they are shown in Fig. \ref{C2m}. As it is clearly
seen from this figure, $C_1$ is always positive and our numerical
 calculation approves this behavior for different temperatures. But $C_2$ becomes
 negative in the near horizon region and according to our results in the previous
 section in this region an imaginary part contributes to the action of the probe brane. Consequently the corresponding mesons in the QGP become unstable. Since the Dilaton field
 is constant in this background, $\eqref{act3}$ and $\eqref{ImS}$ lead to %
\be\begin{split}\label{decay1} %
 \Gamma &\sim \int_0^{\delta r_c} d(\delta r)~ \delta r^3 \sqrt{C_1 \delta r^2+C_2},\cr
 &\sim \frac{2(-C_2)^{\frac{5}{2}}}{15 R_*^4 C_1^2}.
\end{split}\ee %
As it was indicated in the previous section, there is no one to one
correspondence between mass and $R_*$ in the near horizon region.
Hence in order to plot the thermal width in terms of the mass, one
should select the favorable configurations whose $R_*$s are larger
than ${R_*}_{min}$ (see Fig. \ref{mRs}). At fixed temperature, the
behavior of the thermal width with respect to the mass has been
plotted in Fig. \ref{Gm}. We clearly observe that for each value
of the temperature there is a critical value for the mass $m_c$
where for $m>m_c$ the action of the probe brane remains real and as
a result the corresponding mesons in the plasma are stable. In order
to find the value of the critical mass, we set $C_2$ to zero which
leads to a value for $R_*$ named ${R_*}_c$. Solving the equation of
motion for the $R(r)$ with $R_*={R_*}_c$, the asymptotic value of the
resulting ME gives the value of the critical mass. In other words,
the critical value for the mass comes from the minimum value of
$R_*$ for which the action is real. Moreover, the comparison between
plots for different temperatures in Fig. \ref{Gm} shows that the
critical mass increases by raising the temperature as it is
expected. Fig. \ref{Gm} also indicates that for $m<m_c$ the value of
the thermal width increases up to a maximum value $\Gamma_{max}$
which corresponds to the minimum value of the mass, $m_{min}$. The mentioned fact in the literature that the thermal width increases with temperature \cite{CasalderreySolana:2011us} is shown in Fig \ref{Gm} (Right). It is obvious that for a fixed value of mass, for example $m_{min}$, the thermal width corresponding to the higher temperature (green plot) has larger value.

At sufficiently large $m$ (compared with temperature) the MEs are
thermodynamically favorable.  By decreasing the mass, there is a
critical mass $m_c$ at which the action of the probe brane becomes
imaginary and therefore the phase transition between a ME and a BE
may happen. The value of the critical mass is reported as $m_c\sim
0.46 \sqrt{\lambda} T$ (for example see \cite{Herzog:2006gh}). In
our units $\lambda=(2\pi)^2$ and for $T=0.22508 (0.22733)$ the
critical mass is $m_c\sim 0.6505 (0.6570)$. Interestingly, our
result (see Fig. \ref{Gm}) shows that the value for the critical
mass is $0.647947 (0.654426)$ which is in agreement with the
reported result.

\begin{figure}[ht]
\begin{center}
\includegraphics[width=2.6 in]{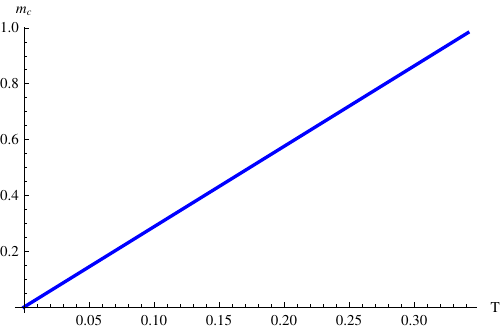}
\caption{Critical mass in terms of temperature. The plot is fitted
with $m_c=2.87875\ T $. } \label{Tmc}
\end{center}
\end{figure}

It is also realized from Fig. \ref{Gm} that the critical mass
varies with the temperature. To see this variation we have plotted
the critical mass in terms of the temperature in Fig. \ref{Tmc}
that clearly shows the linear increase of $m_c$ with respect to the
temperature. Again our result is in agreement with proportionality
between $m_c$ and $T$ reported in the literature \cite{Herzog:2006gh}.

In short, we have introduced a new mechanism for the phase transition between a ME and a BE. As we showed \textit{thermal fluctuations are responsible for the phase transition in the near horizon region}.

Before closing this section, let us discuss a spacial solution for which $R(r)$ is constant. This is the solution when $m\gg T$(or $T=0$). Then, from \eqref{newC}, one can find out that $C_1$ vanishes and $C_2>0$ indicating that the imaginary part of the action disappears. Therefore, the solutions with constant $R(r)$ are stable.

\subsection{Non-zero Magnetic field}
\begin{figure}[ht]
\begin{center}
\includegraphics[width=2.6 in]{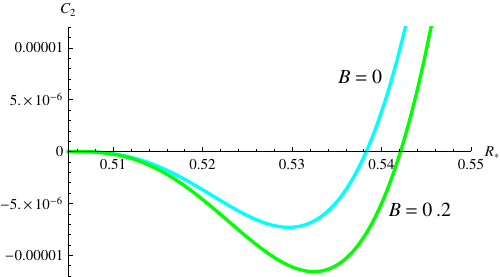}
\includegraphics[width=2.6 in]{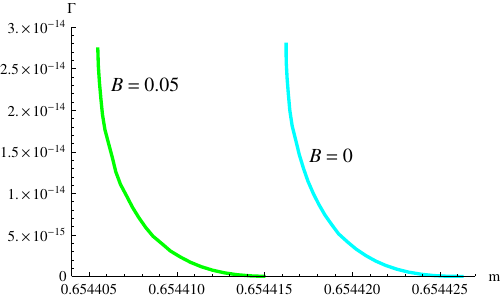}
\caption{Dependence of $C_2$ on $R_*$ (Left) and dependence of
the thermal width on the mass (Right) for two fixed values of the magnetic
field and at the temperature $T=0.22733$.} \label{C2mB}
\end{center}
\end{figure}

In the previous subsection we studied the possibility of the
appearance of thermal width for quarkonia and its dependence on the
mass and the temperature. Now we want to study the same features in
the presence of magnetic field. By turning on the magnetic field,
according to \eqref{MM}, an extra term appears in the definition of
$P$. In the near
horizon region, $C_1$ and $C_2$ are positive and negative,
respectively. Since they are too lengthy, we did not write them here. In Fig. \ref{C2mB}(left), $C_2$ has been plotted in
terms of $R_*$ at a fixed temperature. Although this figure shows
that ${R_*}_c$, for which $C_2$ vanishes, becomes larger in the
presence of magnetic field, the critical mass $m_c$ decreases as it
is shown in Fig. \ref{C2mB}(right) in agreement with the result reported in \cite{Albash:2007bk}. This behavior can be
qualitatively understood from Fig. \ref{RrB}. Similar to the case
of zero magnetic field, the thermal width increases as the mass of
the quarks decreases.

\begin{figure}[ht]
\begin{center}
\includegraphics[width=2.6 in]{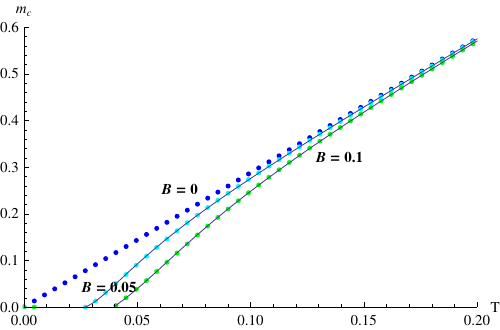}
\includegraphics[width=2.6 in]{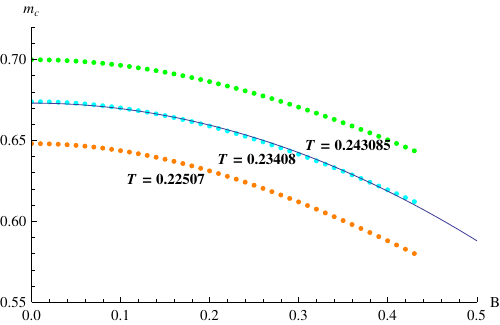}
\caption{Dependence of $m_c$ on the temperature for three values of
the magnetic field(left) and on the magnetic field for three values
of the temperature(right). The blue curves are fitted functions
\eqref{fit} and \eqref{fitmc} for the left and the right plots,
respectively. } \label{TmcB}
\end{center}
\end{figure}

The critical mass as a function of the temperature is plotted in
Fig. \ref{TmcB}(left) for fixed values of magnetic field. Unlike the
case of zero magnetic field, it shows that the critical mass does
not rise linearly with the temperature. The function for critical mass seems to be complicated in the presence of the magnetic field but it may be approximated by the following polynomial
\be\label{fit} %
 m_c=T\sum_{n=0}^5 a_n\left(\frac{(2\pi\alpha')^2 B^2}{T}\right)^n
\ee %
where $a_0=2.87875$ and for $B=0.1$: %
\be\begin{split} %
 a_1=-0.6068, a_2=36.6865, a_3=-719.758, a_4=3202.56, a_5=-4496.51, \nonumber
\end{split}\ee %
and for $B=0.05$:
\be %
 a_1=-1.01318, a_2=206.215, a_3=-12821.7, a_4=151035, a_5=-528361. \nonumber
\ee %
In addition, it turns out from Fig. \ref{TmcB}(left) that at fixed
magnetic field there is a minimum value for the temperature, $T_c$,
where for $T>T_c$ the critical mass is non-zero and a first order
phase transition happens. On the other hand, for $T<T_c<B_z$, the
critical mass vanishes and the first order phase transition
disappears suggesting that MEs are stable configuration in this
region of the temperature. Interestingly, there is a qualitative
agreement between our result and \cite{Albash:2007bk}.

Dependence of the critical mass on the magnetic field is shown in
Fig. \ref{TmcB}(right) for different fixed values of temperature. It
is seen that $m_c$ decreases by increasing $B$ which means that in
the presence of the external magnetic field the probe brane resists
more against the phase transition. From this plot, one may easily
guess that
\be\label{fitmc} %
 m_c=m_0+m_1(2\pi\alpha')^2B^2,
\ee %
where for instance, when the temperature is $T=0.23408$, we obtain $m_c=0.672972$ and $m_1=-0.339937$. Note that there is a maximum value for the magnetic field $B_{max}$ and when $B>B_{max}$ MEs are stable.
It is also obvious from the plot that the starting point of $m_c$ is grater for the larger values of temperature. It means that by increasing temperature the value of the critical mass below which the configuration becomes unstable, increases.

\section{Summary and Conclusion}
In this paper we studied the effect of thermal fluctuations on the
instability of a probe D7-brane, being understood by the appearance
of an imaginary term in action of the brane that may result in
a first order phase transition from Minkowski embedding to
black hole embedding. This study is the gravitational dual of
meson dissociation in the gauge theory. Therefore we corresponded
the imaginary part of the action to the thermal width
of quarkonium mesons.

It was seen that there is a critical mass $m_c$ above it the
thermal width is zero and below it the thermal width increases with
mass until it reaches a maximum at an allowed minimum mass, $m_{min}$.
Furthermore, $m_c$ increases linearly with temperature. As a matter of fact,
for higher temperatures the maximum value of the mass, that below it
quarkonium mesons dissociate, increases. In other words, since $M_{meson}\sim m_q$, it indicates that heavier quarkonia melt in the plasma by increasing the temperature. We also showed that the thermal width increases with temperature for a fixed value of mass.

All the above features had been studied in the presence of the
magnetic field, too, and similar results were obtained. The only
different and notable points in the latter case are a decrease in
$m_c$ with increasing the magnetic field and non-linear relation between
$m_c$ and the temperature. It was seen that in the presence of a non-zero
magnetic field there is a minimum value for the temperature, $T_c$, that $m_c$
is non-zero just for $T>T_c$ and for $T<T_c$ MEs are stable and
accordingly the thermal width is zero.


\end{document}